\documentclass[useASM, referee]{mn2e}
\input{psfig.sty}
\usepackage{graphicx}

\begin{document}

\title{Physical implication of the KRL pulse function of gamma-ray bursts\footnote{send offprint request to: zbzhang@ynao.ac.cn
}}

\date{2004 August 16}
\pubyear{????} \volume{????} \pagerange{2} \onecolumn

\author[Zhang and Qin]
       { Z.-B.\ Zhang$^{1,2}$, Y.-P.\ Qin$^{1,3}$
 \\
        $^1$National Astronomical Observatories/Yunnan Observatory, Chinese Academy of Sciences, \\
            P. O. Box 110, Kunming, Yunnan 650011, P. R. China\\
        $^2$The Graduate School of the Chinese Academy of Sciences\\
        $^3$Physics Department, Guangxi University, Nanning, Guangxi 530004, P. R. China\\
 }

\date{Accepted ????.
      Received ????;
      in original form 2004 November 15}

\pagerange{\pageref{firstpage}--\pageref{lastpage}} \pubyear{2001}

\maketitle

\label{firstpage}

\begin{abstract}
Kocevski, Ryde \& Liang (2003, hereafter Paper I) proposed a
semi-empirical function (the KRL function) of gamma-ray burst
(GRB) pulses, which could well describe those pulses comprising a
fast rise and an exponential decay (FRED) phases. Meanwhile, a
theoretical model which could give rise to this kind of pulse
based on the Doppler effect of the expanding fireball surface was
put froward in details in Qin (2002) and Qin et al. (2004,
hereafter Paper II). To provide a physical explanation to the
parameters of the KRL function, we try to fit light curves of the
Doppler model (Qin 2002; Paper II) with the KRL function so that
parameters in both models can be directly related. We pay our
attention only to single GRB pulses whose profiles are that of
FRED and hence employ the sample presented in Paper I (the KRL
sample) to study this issue. We find from our analysis that, for
light curves, which arise from exponential rise and exponential
decay local pulses, of the Doppler model, the ratio of the rise
index $r$ to the decay index $d$, derived when fitted by the KRL
function, increases quickly first, and then keeps nearly invariant
with the relative width (relative to the timescale of the initial
fireball radius $R_{c}/c$) of local pulses when the width exceeds
2 (the relative width is dimensionless). The rise and decay times
of pulses are found to be related with the Lorentz factor by a
power law, where the power law index associated with the rise time
is less than that of the decay one and both of them are close to
-2. In addition, the mean asymmetry shows a slightly trend of
decreasing with lorentz factors. In plots of decay indexes versus
asymmetry, there is a descending phase and after the phase there
is a rising portion. We find that these long GRBs of the KRL
sample are mainly associated with those light curves arising from
co-moving pulses with the relative width being larger than 0.1.
Shown in our analysis, the effect of the co-moving pulse shape on
the KRL function parameters of the resulting pulses is
considerable and can be distinguished by the decay index $d$ when
the relative co-moving pulse width is less than 2 (when the
relative width is larger than 2, it would be difficult to discern
the difference in the resulting pulse shapes).
\end{abstract}


\section{Introduction}

Light curves of gamma-ray bursts (GRBs) exhibit very complex and
distinct morphologies, without any systematic temporal features
(Fishman \& Meegan 1995). A single spectrum, generally the Band
function spectrum, seems to be a universal characteristic of GRBs
(Piran et al. 1997). The highly variable temporal structure of
GRBs may provide the clue to disclose the puzzle about the
spectrum observed (Sari et al. 1997). It was previously found that
some individual peaks decay more gradually than they rise (e.g.
Fishman et al., 1994; Fenimore E. E., 1999). Many GRBs can well be
decomposed into fundamental pulses (Norris et al. 1996; Lee,
Bloom, \& Petrosian 2000a, 2000b), which comprise a fast rise and
an exponential decay (FRED) phases in general. The pulse-shape
light curves as the elementary events of GRBs probably provide the
intrinsical information. Time asymmetry in GRB light curves has
been discussed by several authors (see Barat et al. 1984; Norris
et al. 1986; Kouveliotou et al. 1992; Link, Epstein, \&
Priedhorsky 1993; Nemiroff et al. 1994). The rise time ($t_r$),
decay time ($t_d$), and full width at half maximum ($FWHM$) are
the main factors concerned (McBreen et al. 2002; Ryde et al. 2003;
Kocevski, Ryde \& Liang 2003, hereafter Paper I). Norris et al.
(1996) utilized a single flexible empirical function to represent
pulses in long bright GRBs, and found the ratio of rise-to-decay
times of pulses to be unity or less. Shown in Ryde \& Svensson
(2000, 2002) are some other fitting profiles.

To fully model the FRED light curves or individual pulse shapes,
Kocevski et al. (2003) (Paper I) put forward an empirical function
(the KRL function) described by
\begin{equation}
F(t)={F_m}(\frac{t}{t_m})^r[\frac{d}{d+r}+\frac{r}{d+r}(\frac{t}{t_m})^{(r+1)}]^{-\frac{r+d}{r+1}},
\end{equation}
where $t_m$ is the time of the maximum energy flux, $F_m$, of the
pulse. The shapes of single GRB pulses are confined by the two
parameters $r$ and $d$. It is a very functional and flexible model
based on the physical first principle and the well-established
empirical correlation between $E_p$ and flux. This function could
be adopted to characterize some GRB pulses such as FRED light
curves. However, what can parameters $r$ and $d$ reveal if the
FRED pulses described by the KRL function arise from fireball
sources?

Recently, a theoretical pulse model (called the Doppler model) was
proposed (see section 2 below) to relate the observed
characteristics of GRB pulses with parameters of the intrinsic
radiation in the framework of fireballs, where the Doppler effect
associated with the expanding fireball surface is the key factor
to be concerned (Qin 2002; Qin et al. 2004, hereafter Paper II).
Illustrated in this model, the profile of an observed pulse would
mainly be determined by the intensity of co-moving pulses,
$\widetilde{I}(\tau_{\theta})$ (see Paper II). It was revealed
that most of the resulting pulses possess a characteristic of
FRED, even though the co-moving pulses concerned are diverse
significantly.

Hinted by the common FRED feature shown in the two models, a
primary goal of this paper is born. We want to know how parameters
$r$ and $d$ would be associated with intrinsic quantities such as
the Lorentz factor $\Gamma$ and the co-moving pulse width
$\Delta\tau_{\theta}$ (see below), once co-moving pulses are
provided (see Paper II). Some prevenient investigations did't take
into account the influence of co-moving pulses on the observed
light curve (see, e.g., Ryde 2004). However, as mentioned above,
it was shown in the Doppler model that the observed pulse shape
(especially its peakedness) is mainly associated with the
co-moving pulse shape and width.

In the following, we will investigate the impacts of exponential
rise and exponential decay co-moving pulses on the observed light
curve in detail. In section 2 we will introduce formulas of the
Doppler model to describe the observed FRED pulses. In section 3,
we will give a reason for choosing the types of co-moving pulse
and will simulate a lot of resulting pulses and then will fit them
with equation (1), and in doing so, some parameters associated
with the observed light curve will be presented. In section 4, we
will contrast our results with those derived from observation. Our
results will be discussed and summarized in section 5.

\section{Formulas of the Doppler model employed in this paper}

We list in this section basic formulas, which were derived and
presented in Paper II, of the Doppler model so that it would be
convenient for us to employ or refer to them in the following
analysis.

Adopting the same symbols assigned and used in Paper II, the count
rate equation derived from the expected flux emitted from an
expanding fireball surface is expressed as (see Paper II)
\begin{equation}
C(\tau)=C_0\frac{\int\limits_{\widetilde{\tau}_{\theta,min}}^{\widetilde{\tau}_{\theta,max}}\widetilde{I}(\tau_\theta)(1+\beta\tau_\theta)^2(1-\tau+\tau_\theta)d\tau_\theta
\int\limits_{\nu_1}^{\nu_2}\frac{g_{0,
\nu}(\nu_{0,\theta})}{\nu}d\nu}{\Gamma^3(1-\beta)^2(1+\frac{\beta}{1-\beta}\tau)^2},
\end{equation}
where $\beta=\sqrt{\Gamma^2-1}/\Gamma$, $\Gamma$ is the Lorentz
factor, $C_0$ is a constant which includes the luminosity distance
$D$ between the fireball and the observer and other factors (see
Paper II), $\widetilde{I}(\tau_\theta)$ represents the development
of the intensity of the co-moving emission (in this paper, the
co-moving pulse), $g_{0,\nu}(\nu_{0,\theta})$ describes the
rest-frame radiation mechanism, and $\nu_{0,\theta}$ is the rest
frame emission frequency which is related with the observed
frequency $\nu$ by the Doppler effect. In equation (2), the count
rate concerned is defined within an energy channel of [$\nu_1$,
$\nu_2$]. As shown in Paper II, letter ``$t$'' denotes an absolute
value of time which is normally defined (it is always in units of
$s$), and the Greek letter ``$\tau$'' represents a relative time
scale (in units of 1) corresponding to ``$t$''. The two types of
quantity are related by (see Paper II)
\begin{equation}
\tau\equiv\frac{t-t_{c}-D/c+R_{c}/c}{R_{c}/c},
\end{equation}
\begin{equation}
\tau_{\theta}\equiv\frac{t_{\theta}-t_{c}}{R_{c}/c},
\end{equation}
\begin{equation}
\tau_{\theta,min}\equiv\frac{t_{\theta,min}-t_{c}}{R_{c}/c},
\end{equation}
and
\begin{equation}
\tau_{\theta,max}\equiv\frac{t_{\theta,max}-t_{c}}{R_{c}/c},
\end{equation}
where $t_{c}$ and $R_{c}$ are constants, and we assign $
t_{\theta,min}\leq t_{\theta} \leq t_{\theta,max}$ (and hence $
\tau_{\theta,min}\leq \tau_{\theta}\leq \tau_{\theta,max}$). The
limits of the integral of $\tau_{\theta}$ in formula (2) are $
\widetilde{\tau}_{\theta,min}=max\{\tau_{\theta,min},(\tau-1+\cos\theta_{max})/(1-\beta\cos\theta_{max})\}$
and $
\widetilde{\tau}_{\theta,max}=min\{\tau_{\theta,max},(\tau-1+\cos\theta_{min})(1-\beta\cos\theta_{min})\}$,
where $\theta_{min}$ and $\theta_{max}$ are the lower and upper
limits of angle $\theta$, respectively (see Paper II). One could
find from these definitions that $\tau$ and $\tau_{\theta}$ are
dimensionless quantities.

Note that, the value of $t_c$ could be arbitrarily chosen (it
depends on the time we assign to that moment). For the sake of
simplicity, we take
\begin{equation}
t_c=(R_c-D)/c
\end{equation}
in the following analysis, where $R_{c}$ is the initial radius of
the fireball (measured at time $t_c$). In this way, one gets
\begin{equation}
\tau=\frac{t}{R_{c}/c}.
\end{equation}
Substituting this relation into formula (2), the count rate as a
function of observation time $t$ could then be plainly
illustrated, so long as $R_{c}$ is available (or assumed).

In this paper, we take the generally adopted spectral form
proposed by Band et al. (1993), the so-called Band function
\begin{equation}
$$
g_{0,\nu,B}(\nu_{0,\theta})=\left\{
\begin{array}{cc}
(\frac{\nu_{0,\theta}}{\nu_{0,p}})^{1+\alpha_{0}}exp[-(2+\alpha_{0})\frac{\nu_{0,\theta}}{\nu_{0,p}}],\ \ & \ \ (\frac{\nu_{0,\theta}}{\nu_{0,p}}\leq\frac{\alpha_{0}-\beta_{0}}{2+\alpha_{0}})\\
(\frac{\alpha_{0}-\beta_{0}}{2+\alpha_{0}})^{\alpha_{0}-\beta_{0}}exp(\beta_{0}-\alpha_{0})(\frac{\nu_{0,\theta}}{\nu_{0,p}})^{1+\beta_{0}}\ \ & \ \ ({\rm } \frac{\nu_{0,\theta}}{\nu_{0,p}}>\frac{\alpha_{0}-\beta_{0}}{2+\alpha_{0}})\\
\end{array}
\right.
$$,
\end{equation}
as the rest frame spectrum $g_{0,\nu}(\nu_{0,\theta})$.

One might observe that the profile of the KRL function are
determined by parameters $r$ and $d$. However, in the Doppler
model, the characteristics of an observed pulse would be obviously
influenced by the Lorentz factor and the co-moving intensity of
radiation (see Paper II). Thus, by fitting light curves of formula
(2) with function (1), one might be able to tell how parameters
$r$ and $d$ and various widths of the KRL function are related
with the Lorentz factor and co-moving pulses, assuming that the
sources observed are undergoing a fireball stage.

\section{General analysis on the pulses of fireball sources}

In this section, we study the light curve of formula (2) in a
particular case: the adopted co-moving pulse is an exponential
rise and an exponential decay one. Various sets of parameters $r$
and $d$ of the KRL function would be obtained when fitting light
curves of formula (2) associated with different Lorentz factors
and different widths of the co-moving pulse with equation (1). In
this process, we consider the radiation from the whole spherical
surface of the fireball, although the contribution from the area
of $\theta>1/\Gamma$ is insignificant. (Impacts of different
emission areas in the fireball surface on the resulting pulses
were shown in Paper II.) Thus, we adopt $ \theta_{min}=0$ and
$\theta_{max}=\pi/2$. Not losing the generality, we assign
$C_0=1$. In addition, we take $ \nu_{1}=50\nu_{0,p}$ and
$\nu_{2}=100\nu_{0,p}$, where $\nu_{0,p}$ is the break frequency
of the rest frame Band function spectrum, so that when assigning
$\nu_{0,p}=1keVh^{-1}$ the energy range would correspond to the
second channel of BATSE.

It is supposed that a very violent collision of two shells gives
rise to the observed GRB pulse behavior, and the increase (or
decrease) of co-moving emission is proportional to the total
radiation in the rise (or decay) phase of the co-moving pulse,
namely
\begin{equation}
$$
dI(\tau_{\theta})=\left\{
\begin{array}{cc}
\frac{1}{\sigma_{r}}I(\tau_{\theta})d\tau_{\theta} \ & \ \
(in\ the\ rise\ phase)\\
-\frac{1}{\sigma_{d}}I(\tau_{\theta})d\tau_{\theta} \ & \ \
(in\ the\ decay\ phase)\\
\end{array}
\right.
$$,
\end{equation}
where $\sigma_{r}$ and $\sigma_{d}$ are two positive constants.
Integrating this equation leads to
\begin{equation}
$$
I(\tau_\theta)=I_0\left\{
\begin{array}{cc}
exp[(\tau_{\theta}-\tau_{\theta,0})/\sigma_{r}] \ & \ \ (\tau_{\theta,min}\leq\tau_{\theta}\leq\tau_{\theta,0})\\
exp[-(\tau_{\theta}-\tau_{\theta,0})/\sigma_{d}] \ & \ \ (\tau_{\theta,0}<\tau_{\theta}\leq\tau_{\theta,max})\\
\end{array}
\right.
$$,
\end{equation}
where $I_0$ and $\tau_{\theta,0}$ are the integral constants.
Associated with the well-known mechanisms, the rising part of this
co-moving pulse might connect to the shell crossing time and the
decay portion might relate to the cooling time. (It is interesting
that a pulse form with an exponential rise and an exponential
decay phases was previously adopted as an empirical function to
describe some observed GRB pulses by Norris et al. 1996). The
spiky form of this co-moving pulse implies that the physical
interaction processes of shells are greatly impetuous and rapid.

In the process of the shell collision, two intrinsic timescales,
the cooling timescale of electrons and the shell crossing
timescale, should be considered simultaneously, and they together
would cause the width of the co-moving pulse. In terms of the
Doppler model, co-moving pulses would be significantly altered by
the expanding fireball surface and then would lead to the observed
forms of pulses (see Paper II). An alternative interpretation to
this is that the two timescales together with the curvature
timescale which is due to the relativistic kinematics of expanding
shell would give rise to the observed pulses. In many cases, the
curvature time and the shell crossing time dominate over the
cooling time (See, Ryde et al. 2003). Spada et al. (2000)
suggested that this case will occur at a distance
$R<5\times10^{14}cm$, but for larger radii the radiative cooling
time will become the dominant contribution to the pulse duration
(in the first case the cooling timescale would be relatively small
while in the last case it would be relatively large).

For convenience, the width of co-moving pulse (11) is defined as
\begin{equation}
\Delta\tau_{\theta}=\tau_{\theta,max}-\tau_{\theta,min}.
\end{equation}
Note that the quantity $\Delta\tau_{\theta}$ determined by eqs.
(5) and (6) is also dimensionless. In the following analysis we
assign $\tau_{\theta,min}=-\sigma_{r}$,
$\tau_{\theta,max}=\sigma_{d}$ and $\tau_{\theta,0}=0$ in order to
check whether the ratio of the rise time to the decay time of
co-moving pulses can influence the observed light curves. Here we
consider two kinds of spiky co-moving pulses. One concerns
$\sigma_r=\sigma_d$ (case 1) and the other is associated with
$\sigma_r=4\sigma_d$ (case 2).

\subsection{In the case of $\sigma_{r}=\sigma_{d}$}

Here, we consider co-moving pulse (11) with
$\sigma_{r}=\sigma_{d}$, which implies that the cooling time and
the shell crossing time are comparable.

\subsubsection{Impact of the co-moving pulse width}

To investigate how parameters $r$ and $d$ are related with the
co-moving pulse width, we calculate various light curves of
formula (2), arising from co-moving pulse (11) with
$\Delta\tau_{\theta}$ = 0.001, 0.002, 0.005, 0.01, 0.02, 0.05,
0.1, 0.2, 0.5, 1, 2, 5, 10 respectively (The reason to choose
these values is discussed in Appendix B.), and then fit them with
the KRL function. For the rest frame Band function spectrum we
take $\alpha_{0}=-1$ and $\beta_{0}=-2.25$, and for the Lorentz
factor we consider $\Gamma$ = 2, 5, 10, 100, 1000, and 10000.
Parameters $r$ and $d$ are directly obtained from the fit. The
rising width $t_r$ and the decaying width $t_d$ of the observed
pulse would be obtained from equation (1) when the fitting
parameters are applied. (Note that, quantities $t_r$ and $t_d$ are
consistent with those denoted respectively in Paper I.) We define
the asymmetry of the observed pulse as $t_r/t_d$, the ratio of the
rise fraction timescale of the pulse to the decay fraction
timescale.

(i) Relations between these parameters ($r$, $d$, $r/d$, $t_r$,
$t_d$, $t_r/t_d$) and the co-moving pulse width
$\Delta\tau_{\theta}$ are shown in Fig. 1. Shown in panel (a),
there lie minimums of $d$ between $\Delta\tau_{\theta}=0.03$ and
0.1. From panels (a), (b), (c) and (f), we find that $d$, $r$,
$r/d$ and $t_r/t_d$ are independent of $\Delta\tau_{\theta}$ when
$\Delta\tau_{\theta}$ is larger than 2. In terms of the KRL
function, the result suggests that profiles of the observed light
curves would not be distinguishable when the width of co-moving
pulses is large enough (say $\Delta\tau_{\theta}>2$), which is in
agreement with what revealed in Paper II (see Paper II Fig. 3).
According to equation (4), a large value of $\Delta\tau_{\theta}$
suggests a relatively small value of the radius of the fireball,
and this in turn indicates that the curvature timescale is
relatively small. In this situation, it would be reasonable when
the cooling time plus the shell-crossing time dominate far over
the curvature timescale. However, as shown in panel (f), the pulse
asymmetry would decease rapidly with the decreasing of the
co-moving pulse width when $\Delta\tau_{\theta}<2$. Panels (d) and
(e) indicate that $t_r$ increases with the increasing of
$\Delta\tau_{\theta}$ at all time, while $t_d$ is sensitive to
$\Delta\tau_{\theta}$ only when $\Delta\tau_{\theta}$ is large
enough (say $\Delta\tau_{\theta}>0.1$). This must be due to the
fact that when the curvature timescale dominates far over the two
other timescales, the decay phase of the light curve would be
determined by the curvature effect and its profile would remain
unchanged when $\Delta\tau_{\theta}<0.1$(approaching the so-called
standard form defined in Paper II; see also Paper II Fig. 3). A
contrast between panels (c) and (f) suggests that the ratio of $r$
to $d$ doesn't adapt to characterize the asymmetry of the observed
pulses.

\begin{figure}
\centering
 \includegraphics[width=5.4in,angle=0]{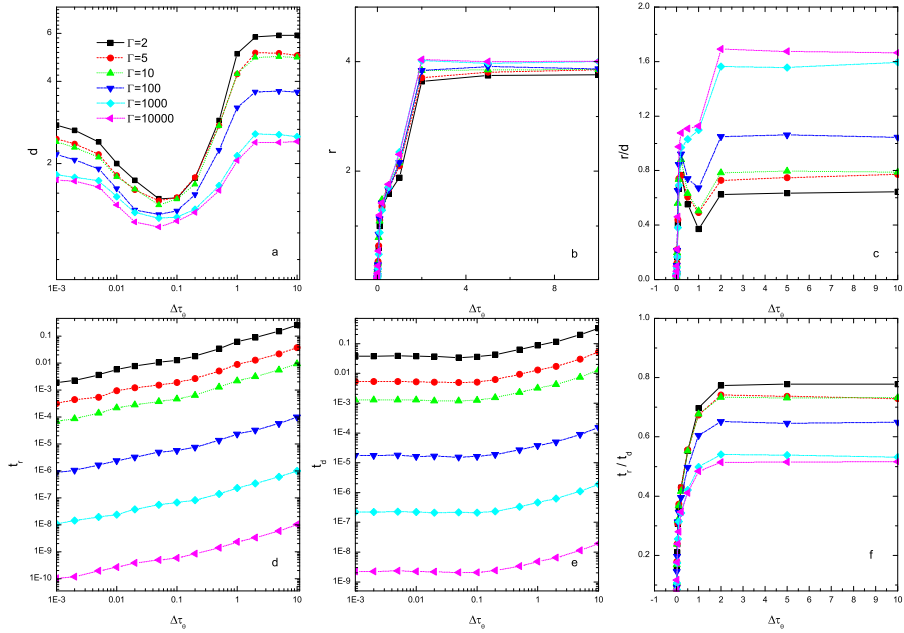}
    \caption{Relations of parameters ($d$, $r$, $t_r$, $t_d$, $r/d$, and $t_r/t_d$) of the observed light curves and
the co-moving pulse width ($\Delta\tau_{\theta}$) for different
lorentz factors ($\Gamma$ = 2, 5, 10, 100, 1000, 10000)
respectively. Symbols corresponding to different values of
$\Gamma$ are displayed in panel (a).}
  \label{fig1}
 \end{figure}

(ii) Displayed in Fig. 2 are the relations between parameters $d$,
$r$, $r/d$, $t_r/t_d$, $FWHM$, and $t_r/FWHM$. Shown in panel (a),
the smallest value of index $d$ could be detected in the range of
$0.5<r<1$. Panels (b) and (f) also demonstrate the existence of a
minimum of $d$, with the minimum corresponding to smaller values
of $t_r/t_d$ and $t_r/FWHM$ when the Lorentz factor becomes
larger. One finds from panel (e) that, for a given Lorentz factor,
the pulse asymmetry is very sensitive to the co-moving pulse width
when the latter is very small, and when the latter becomes large
enough, the asymmetry would become invariant. This is in good
agreement with what illustrated above. In addition, panel (e)
shows that the larger the Lorentz factor the narrower the pulse
observed, which is in consistent with what previously known (see,
e.g., Fenimore et al. 1993).

\begin{figure}
\centering
  \includegraphics[width=5.4in,angle=0]{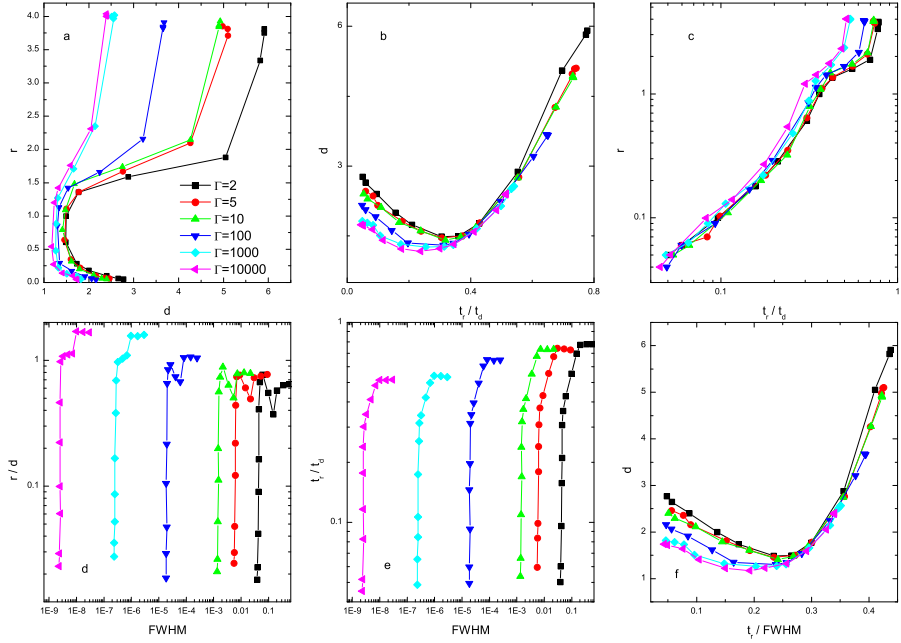}
   \caption{Relations between parameters $d$, $r$, $r/d$, $t_r/t_d$,
$FWHM$, and $t_r/FWHM$, where we adopt $\Gamma$ = 2, 5, 10, 100,
1000, 10000 respectively. Symbols associated with different
Lorentz factors are denoted in panel (a).}
  \label{fig2}
 \end{figure}

We find that characteristics of the relationships displayed in
Figs. 1 and 2 are the same for different Lorentz factors.

\subsubsection{Impact of the Lorentz factor}

Here, we study how parameters $d$, $r$, $r/d$, $t_r$, $t_d$ and
$t_r/t_d$ are related with $\Gamma$ when different values of
$\Delta\tau_{\theta}$ are adopted. We take $\Gamma$ = 2, 5, 10,
70, 100, 150, 200, 500, 1000, 10000 respectively. Five typical
values of the co-moving pulse widths, $\Delta\tau_\theta$ = 0.001,
0.01, 0.1, 1, and 10, are adopted. The indexes of the rest frame
Band function spectrum are the same as those adopted above. In the
same way, light curves of (2) associated with various sets of
these intrinsic parameters are calculated and then are fitted with
equation (1).

\begin{figure}
\centering
\includegraphics[width=5.4in,angle=0]{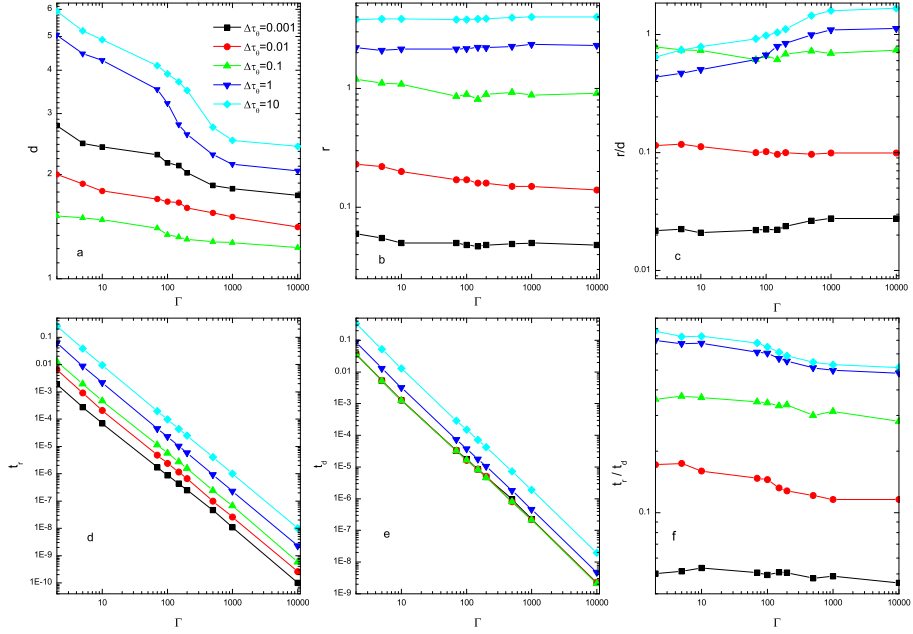}
    \caption{Relations of parameters ($d$, $r$, $t_r$, $t_d$, $r/d$,
   $t_r/t_d$) of the observed light curves and the Lorentz factor $\Gamma$ associated with different values of
   the co-moving pulse width ($\Delta\tau_{\theta}$ = 0.001, 0.01, 0.1, 1, 10)
   respectively. Symbols are denoted in panel (a).}
  \label{fig3}
 \end{figure}

Shown in Fig. 3 are the relations of parameters of the KRL
function associated with the observed light curves and the Lorentz
factor. As shown in panel (f), the effects of $\Gamma$ on the
pulse asymmetry are negligible, although a weak anti-correlation
between $t_r/t_d$ and $\Gamma$ is visible. The following relations
could be concluded from panels (d) and (e):
\begin{equation}
t_r\propto\Gamma^{\alpha_r}
\end{equation}
and
\begin{equation}
t_d\propto\Gamma^{\alpha_d}.
\end{equation}
Values of $\alpha_r$ and $\alpha_d$ could be obtained by
performing a linear fit to the logarithmic format data of the two
panels. The results are listed in Table 1. The relation
$\Delta\tau_{FWHM}\propto \Gamma^{-2}$ was obtained by Qin et al.
(2004) (Paper II) when an extremely narrow co-moving pulse is
concerned. The index of $-2$ is nothing but merely a result of
time compression effect caused by the forward motion of the ejecta
(see Appendix A). The values of $\alpha_r$ and $\alpha_d$
presented in Table 1 are close to $-2$, which must be due to the
same time compression effect. Note that the observed timescale
lasts a little longer (as the indexes are slightly larger than
$-2$) than what the time compression effect suggests. This could
be understood when one recalls that what we consider in this paper
is a fireball surface rather than an ejecta moving towards the
observer (in the latter case, $\theta=0$) and the emission of the
surface lasts an interval of time (if the emission is extremely
short, we come to Paper II equation [44]). Shown in Table 1 we
observe that $\alpha_d$ is slightly larger than $\alpha_r$. This
suggests that the rising part of the observed pulse is less
affected by the fireball surface than the decay portion is (the
less affected by the fireball surface the closer the index to
$-2$).

\begin{table*}
\centering \caption{Power law indices $\alpha_r$ and $\alpha_d$
derived for cases 1 and 2} \label{Tab.1}
\begin{tabular}{|c|c|c|c|c|c|}
\hline\hline
case &$\Delta\tau_{\theta}$&$\alpha_r$&$\chi^{2}/\nu$& $\alpha_d$&$\chi^{2}/\nu$\\
\hline
    &0.001& -1.95 &2.7&-1.92&3.1\\
     &0.01&-1.98 &3.0&-1.93&2.6\\
$\sigma_r=\sigma_d$ &0.1&-1.97 &2.7&-1.93&2.9\\
     &1&-1.99 &3.2&-1.94&2.8\\
     &10&-1.99&3.0&-1.94&2.7\\
\hline
&0.001& -1.90 &2.7&-1.87&2.8\\
     &0.01& -1.91&2.9&-1.87&2.7\\
$\sigma_r=4\sigma_d$ &0.1&-1.90 &3.0&-1.87&2.8\\
     &1&-1.94 &2.9&-1.89&3.0\\
     &10&-1.93&3.0&-1.88&3.0\\
\hline
\end{tabular}
\end{table*}

(ii) Relations between parameters shown in Fig. 2, the characters
of the observed light curves (ie., $d$, $r$, $r/d$, $t_r/t_d$,
$FWHM$, and $t_r/FWHM$), are displayed in Fig. 4, where different
data points correspond to different values of $\Gamma$ (data
points associated with the same value of $\Delta\tau_{\theta}$ are
denoted by the same symbol).

\begin{figure}
\centering
  \includegraphics[width=5.4in,angle=0]{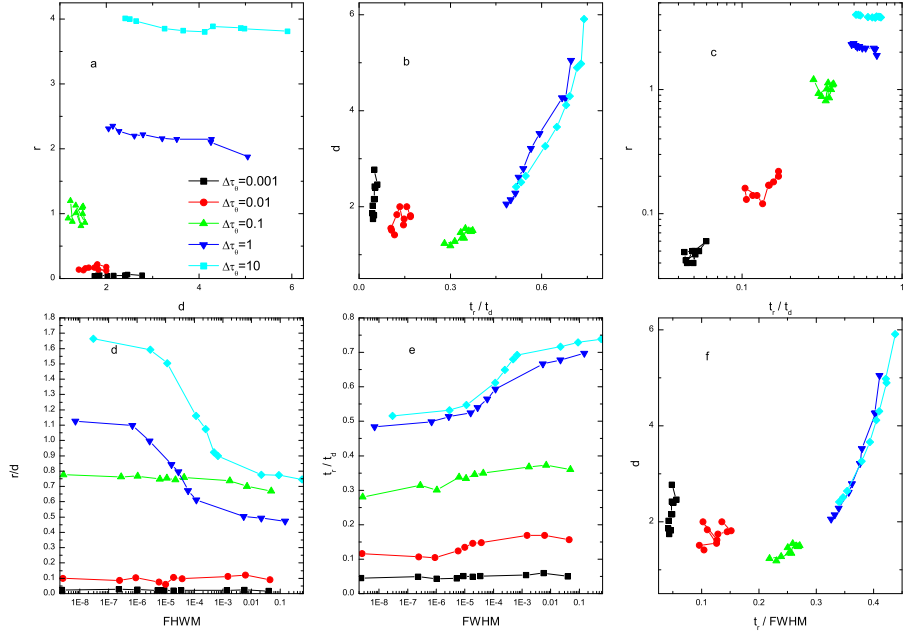}
   \caption{Relations between parameters of the observed light curves associated
with $\Delta\tau_{\theta}$ = 0.001, 0.01, 0.1, 1, 10 respectively.
Symbol are denoted in
   panel (a).}
  \label{fig4}
 \end{figure}

Panel (e) shows that, for very narrow co-moving pulses, asymmetry
of the observed pulses would keep to be invariant with $FWHM$,
while for wider co-moving pulses, the asymmetry would decrease
first and then would keep to be invariant with the decreasing of
$FWHM$ (here, since the co-moving pulse width is fixed, the
decreasing of $FWHM$ would be caused by the increasing of the
Lorentz factor). Panels (b), (c) and (f) are similar to the
corresponding panels in Fig. 2. They implies that: (1) there is a
correlation between parameters $d$ and $t_r/FWHM$ in
$t_r/FWHM\geq0.2$ (or $\Delta\tau_\theta\geq0.1$) and an
anti-correlation between the two quantities in $t_r/FWHM\leq0.2$
(or $\Delta\tau_\theta\leq0.1$) (see section 4 for a detailed
discussion), no matter what the Lorentz factors are. (2) a power
law relation probably exists between parameters $r$ and $t_r/t_d$.
Let us assume
\begin{equation}
t_r/t_d\propto r^\epsilon.
\end{equation}
Fitting the logarithmic format data with a linear function, we
obtain $\epsilon=0.578$. It shows that index $r$ and the pulse
asymmetry tie in, and the former can measure the latter as long as
the co-moving pulse is known.


The result that the pulse asymmetry for long bursts will decrease
as the $FWHM$ narrows supports previous conclusions found by
Norris et al. (1996) (see also e.g. Reichart et al. 2001). As
shown in Paper II, a small value of $FWHM$ could be caused by a
small co-moving pulse width, or a small fireball radius, or a
large Lorentz factor. Of the three factors, the third is the most
sensitive one according to the Doppler model.

\subsection{In the case of $\sigma_{r}=4\sigma_{d}$}

Now we study the case of adopting co-moving pulse (11) with
$\sigma_{r}=4\sigma_{d}$ (case 2), which corresponds to a
relatively fast cooling timescale of electrons.

In order to find out whether there is any difference between the
two cases, we perform the same analysis as that in case 1. The
only difference in this analysis is that we replace
$\sigma_{r}=\sigma_{d}$ with $\sigma_{r}=4\sigma_{d}$ for
co-moving pulse (11). It is surprised that characteristics shown
in the corresponding relations in the two cases are very similar.
Conclusions drawn from case 1 holds in case 2. However, parameter
$d$ shows a difference in the two cases, which is illustrated in
Fig. 5 (presented in this figure is also the comparison between
the values of $r$ for the two cases). To produce the data of this
figure, we take $\Gamma=100$. As suggested by the figure, index
$d$ is more sensitive to the co-moving pulse than $r$ is. This
enables us to relate an observed pulse with the corresponding
co-moving pulse by parameter $d$ so long as $\Delta\tau_\theta$ is
smaller than 2. It suggests that a lager value of $d$ may
correspond to a faster cooling process. Once $\Delta\tau_\theta$
becomes larger than 2, it would be difficult to discern different
co-moving pulses from $d$.

\begin{figure}
\centering
   \includegraphics[width=4.in,angle=0]{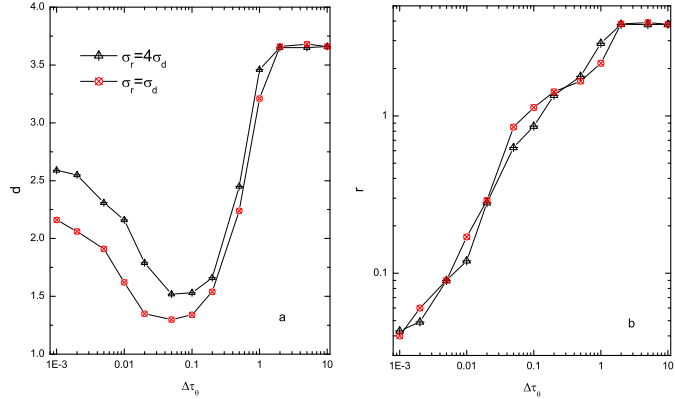}
  \caption{Relations between parameters ($d$ and $r$) of the observed light curve and the co-moving pulse width
  for cases 1 and 2. Here we take $\Gamma$ = 100. Relations associated with $d$ and $r$ are presented in panels (a) and (b), respectively.
  Symbols are denoted in panel (a).}
  \label{Fig5}
\end{figure}

The analysis shows that parameters $\alpha_r$ and $\alpha_d$ in
case 2 are larger than those in case 1. This indicates that the
influence of the co-moving pulse in case 2 on observations is
greater than that in case 1.

In case 2, regarding relation (15) we get $\epsilon=0.464$ by a
linear fit. It shows that the same conclusion obtained in case 1
holds in this case.

\section{Comparison with the observed data}

Let us contrast Fig. 4 panel (f) with Paper I Fig. 12 to provide a
direct comparison between the observed data and the expectation of
the Doppler model in the relationship between $d$ and $t_r/FWHM$.
In Fig. 4 panel (f), data in the decaying portion of the
relationship curve (the data associated with
$\Delta\tau_{\theta}<0.1$) are defined as sample 1, while those in
the rising portion (the data associated with
$\Delta\tau_{\theta}\geq0.1$) are called sample 2. The
corresponding sets of data in case 2 are also called sample 1 (the
data associated with $\Delta\tau_{\theta}<0.1$) and sample 2 (the
data associated with $\Delta\tau_{\theta}\geq0.1$) respectively.
The data presented in Paper I include 77 individual pulses with
time profiles longer than 2 seconds, which we define as sample 3.

One might observe that analysis performed in the previous sections
is based on the concept of $\tau$ which is dimensionless. Relation
between this quantity and the observed time $t$ is shown in
equation (8). As $\tau$ is proportional to $t$, parameter
$t_r/FWHM$ would be the same in both definitions. Fortunately, $d$
is dimensionless. We therefore can directly compare data of
samples 1 and 2 with those of sample 3. Plots of $d$ vs.
$t_r/FWHM$ for cases 1 and 2 together with the data of sample 3
are shown in Fig. 6 panels (a) and (b) respectively. Presented in
the figure is also the division (at $t_r/FWHM\simeq0.2$) between
samples 1 and 2.

\begin{figure}
\centering
   \includegraphics[width=5.2in,angle=0]{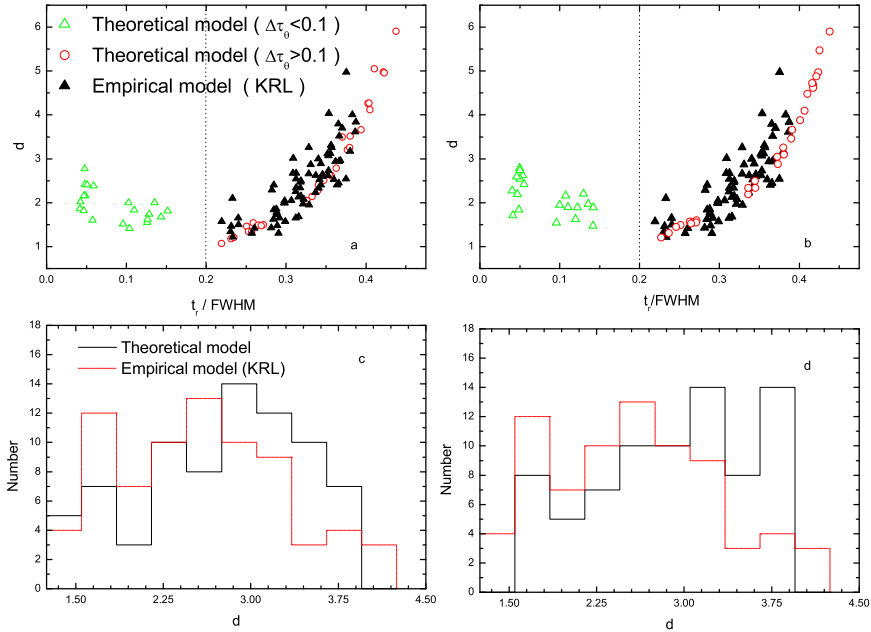}
  \caption{Comparison between sample 3 and samples 1 and 2 in the planes of $d$ vs. $t_r/FWHM$ and number vs. $d$,
  where 1) panel (a) is associated with case 1 while panel (b) corresponds to case 2.
  [Note that, data of samples 1 and 2 plotted in panel (a) are merely the replicate of Fig. 4 panel (f).]
  The vertical line denotes the position of $t_r/FWHM\simeq0.2$
  (or $\Delta\tau_\theta\simeq0.1$) which divides data of sample 1 from
  those of sample 2. Symbols are denoted in panel (a). 2) Panels (c) and (d) shows distributions of index $d$ for
  case 1 and case 2 respectively. The assumed lognornmal distribution of $\Delta\tau_{\theta}$
  [see eq.(16)]
  yields the theoretical $d$ distribution as the lorentz factor
  is taken one typical value, say, $\Gamma=100$. Same symbol meanings as in panel(c).}
  \label{Fig6}
\end{figure}


We find in surprise that sample 3 is well within the range of
sample 2, while it is completely irrelevant to sample 1. In
addition, sample 3 shows a positive correlation between $d$ and
$t_r/FWHM$ just as what sample 2 shows. To investigate whether
samples 2 and 3 have indeed the same distribution on a certain
significance level, we take a general K-S test to the
two-dimensional distributions of the two samples. We adopt the
effective D definition of two-dimensional K-S statistic D as the
average of two values obtained by above samples individually.(see,
Press et al.(1992)) The K-S statistic and the indicated
probability (significance level) are listed in Table 2.
\begin{table*}
\centering \caption{Parameters gained by two- and one-dimensional
K-S test for respective cases 1 and 2 } \label{Tab.3}
\begin{tabular}{|c|c|c|c|}
\hline\hline  Panels & KS & Probability & Degrees of freedom\\
\hline
 a &0.3075  &   0.0475&n1=30, n2=76 \\
\hline
b &0.4145  &  0.0022&n1=30, n2=76\\
\hline
c&0.2237 &   0.0376&n1=n2=76 \\
\hline
d &0.2895 &   0.0026&n1=n2=76 \\
\hline
\end{tabular}
\end{table*}
Above results in both case 1 and case 2 imply that sample 2 and
sample 3 are not significantly different in terms of statistics.
Based on the fact that sample 3 is composed of 77 individual
pulses with durations longer than 2 seconds, we thus deduce that
long GRBs could be mainly superposed by wider rest-frame radiation
pulses with $\Delta\tau_\theta\geq0.1$. Motivated by the viewpoint
that the short bursts with $T_{90}<2.6s$ have different temporal
behaviors compared with the long ones and they may actually
constitute a different class of GRBs (e.g. Norris, Scargle, \&
Bonnell 2001), we suppose sample 1 with $\Delta\tau_\theta<0.1$
might hold the characteristic of short GRBs. The fact that sample
2 associated with $\Delta\tau_\theta\geq0.1$ is consistent with
sample 3 in distributions seems to demonstrate at least some of
sources in sample 3 could be described by Qin's theoretical model
in a sense of practice.

To check if the distribution of index $d$ in sample 3 is indeed
expectable, we plot the distribution of $d$ that is found from
Fig. 1a for a assumed distribution of $\Delta\tau_\theta$. The
rise times ($t_{r}$), fall times ($t_{d}$), FWHM, as well as the
time intervals between pulses had been measured and found to be
consistent with lognormal distributions for both short and long
GRBs \cite{MQ02}. Now, let us also consider Gaussian distribution
of the logarithmic format of quantity $\Delta\tau_\theta$:
\begin{equation}
p(log \Delta\tau_\theta)\propto\exp[-\frac{1}{2}(\frac{log
(\Delta\tau_\theta)+0.15}{0.85})^{2}],
\end{equation}
So, many random values of $\Delta\tau_{\theta}$ could be yielded
so long as the reasonable range of $\Delta\tau_{\theta}$ can well
be determined. In Fig. 1a, suppose we take $\Gamma=100$, a
correlation between $d$ and $\Delta\tau_{\theta}$ within the range
of $0.1\leq\Delta\tau_{\theta}\leq2$ (Note that, resulting pulses
within this range are correlated with sample 3 and can then be
distinguished with parameter $d$ in KRL function) offers us an
opportunity to gain the index $d$ corresponding with
$\Delta\tau_{\theta}$. Plots of distribution of $d$ for cases 1
and 2 together with the data of sample 3 are shown in Fig. 6
panels (c) and (d) respectively. Likewise, the K-S test to above
distributions is also made to give the statistic and indicated
probability listed in Table 2. We find the two distributions are
not significantly different from viewpoint of statistics.

\section{Conclusions and discussions}

According to above-mentioned analyses, we can draw the underlying
conclusions:

First, effects of different sorts of spikier co-moving pulses on
the observed pulse shapes could be distinguished by shape
parameter $d$ provided $\Delta\tau_\theta$ is less than 2 or so.
However, parameter $r$ in KRL function and pulse asymmetry could
be bound with the relationship $t_r/t_d \propto r^{\epsilon}$.
Secondly, the asymmetry of observed pulses exhibits a slight trend
of decrease with the increasing of $\Gamma$, whereas it increases
quickly with the increasing of $\Delta\tau_\theta$ when
$\Delta\tau_\theta\leq 2$, beyond this range, it will keep
invariant with $\Delta\tau_\theta$. In observer framework, we find
the asymmetry increase quickly first with $FWHM$ and then behaves
nearly independent of $FWHM$ for an assumed $\Gamma$ or a lager
$\Delta\tau_{\theta}$. Thirdly, two power law relations,
$t_r\sim\Gamma^{\alpha_{r}}$ and $t_d\sim\Gamma^{\alpha_{d}}$ have
been surprisingly found to show the width of observed pulse is
highly sensitive to lorentz factors. Along with previous result
(Paper II eq.[4]), we attribute these properties to the so-called
time compression effect, which is purely kinematical and
independent of physical mechanism, as long as the emission comes
from an expanding fireball. Further, the difference of indexes
$\alpha_{r}$ and $\alpha_{d}$ in case 1 from those in case 2
suggests that diverse intrinsic emission processes may cause
distinct influences on observed profiles.

Following from the discussion in $\S 4$, for reasons not presently
understood, the result that there is a decay portion (sample 1) in
plot of $d$ vs. $t_{r}/FWHM$ which disappears in Kocevski's plot
(Paper I, Fig. 12) is rather surprising. We believe that there
might be some bursts arising from short co-moving pulses
associated with $\Delta\tau_\theta<0.1$ and their corresponding
pulses would be observed if our sample is large enough, and in
this case the corresponding data would be located within the
descending portion in this plot. The two sub-classes of GRBs show
completely different behaviors though they can be explained with
the same emission mechanism, which in general being ascribed to
synchro-Compton radiation via internal shocks, not the external
shocks (see, Piran et al. 1997, Nakar et al. 2001 \& McBreen et
al. 2001, 2002). Based on this comprehension, we infer that the
decay portion probably encompasses the short bursts or at least
some of them.

As mentioned above, our whole investigations are built on the
assumptions of the fireball-model and the isotropic radiation.
Further, all the co-moving pulses concerned in this paper have a
certain width, that is $\Delta\tau_\theta>0$. Previous studies on
light curves of a co-moving $\delta$-function pulse (see, Paper
II) found that this very narrow pulse will lead to a standard form
light curve, which has only decay phase in the resulting pulses
due to pure curvature effect. As it shows in Fig. 1 that the
difference in contribution of $\Delta\tau_\theta$ to rise times
and decay times of light curves is visible. This shows the rise
phase of pulses as a result of the contribution of
$\Delta\tau_\theta$ would reflect not only the energizing of the
shell but also the radiative cooling of electrons, while the decay
portion of the observed pulse could be fully characterized by all
above-mentioned timescales (namely, curvature time ,
shell-crossing time and radiative cooling time). In the case of
$\Delta\tau_\theta\gg0.1$, the Doppler model is not the major
contributor of the pulse shape and indeed the co-moving behavior
could be important. The conclusion is just in excellent agreement
with that of Spada et al.(2000)

By simulating many resulting pulses with several sorts of
co-moving pulses, We find the resulting pulses modelled by Qin's
model are mainly determined by their corresponding co-moving forms
and lorentz factors. Assuming one monotonic rise (or decay)
function is selected to stand for the co-moving form, the
resulting pulses will show concave (or convex) in phase of rise.
In addition, we find the kurtosis especially the rise part of
resulting pulses mainly originates from the spiky co-moving forms
as long as $\Delta\tau_\theta$ is not very narrow enough,
otherwise, these peaked resulting pulses will behave greatly
similar to the so-called standard forms. In other cases, we can
always achieve the flat-peaked resulting pulses provided that the
decay phase of co-moving pulses is in existence.

From an viewpoint of observation, how to choose the co-moving
pulse form for an observed GRB is an urgent issue. Within the
range of $0.001\leq\Delta\tau_{\theta}<2$ (namely, 0.5 s
$\leq$``t'' $\leq$1000 s), the shape of resulting pulses could be
well distinguished by shape-related parameter $d$. This may
enables us to regard $d$ as a probe to speculate on what the
detailed co-moving forms are. On the other hand, we also need
synthetically take into account the special physical process.
Motivated by these considerations, we might give the relatively
correct co-moving pulse form which is then utilized to fit to
observed data of GRBs. With the best fit-of-goodness, the likely
parameters in rest-frame such as $\Gamma$, $\Delta\tau_{\theta}$
and $R_{c}$ could be derived. Once the co-moving pulse shape is
adopted, it can be applied not only to derive some parameters in
co-moving framework, but also to constrain physical emission
mechanism and improving other theoretical models. In this paper,
we primarily focus on the theoretic analysis of pulses. Some
results could be primary and need to be approved by more
observations.

\section*{Acknowledgments}

This work was supported by the Special Funds for Major State Basic
Research Projects (973) and National Natural Science Foundation of
China (No. 10273019). We would like to thank the anonymous referee
for suggestions that were very useful in preparing this manuscript
for publication.

\appendix

\section[A]{Time compression effect}

Here we show how the index of the power law relation between an
observed timescale and the Lorentz factor is $-2$.

Let us consider an ejecta moving towards the observer with a
velocity of $v=c\beta$, which emits two photons at different
times. Suppose the ejecta emits the first photon from distant $D$
at its co-moving time $t_{\theta,1}$, and emits the second one
from distant $D-v(t_{\theta,2}-t_{\theta,1})$ at time
$t_{\theta,2}$. It is obvious that the observer receives the two
photons at its observed time $t_{1}=t_{\theta,1}+D/c$ and
$t_{2}=t_{\theta,2}+[D-v(t_{\theta,2}-t_{\theta,1})]/c$,
respectively. This leads to
\begin{equation}
t_{2}-t_{1}=t_{\theta,2}-t_{\theta,1}-\beta(t_{\theta,2}-t_{\theta,1}).
\end{equation}
Applying $1-\beta\simeq1/2\Gamma^2$ one gets
\begin{equation}
t_{2}-t_{1}\simeq\frac{t_{\theta,2}-t_{\theta,1}}{2\Gamma^2}.
\end{equation}

\section[B]{Discussion about taking the value of co-moving pulse width($\Delta\tau_{\theta}$)}

Since the fireball becomes optically thin when radiation of
$\gamma$-ray begins, the size of it can be estimated, in general,
to be $10^{13-17}$cm (Piran, 1999). Thus
\begin{equation}
R_c/c\sim 10^{3-7}sec
\end{equation}
From (4), we can get
\begin{equation}
\Delta\tau_{\theta}=\frac{\Delta t_{\theta}}{R_c/c}
\end{equation}
By using another relation
\begin{equation}
\Delta t_\theta=\frac{\Delta t}{1-\beta\cos\theta}
\end{equation}
\cite{Qi04} \\
we can deduce
\begin{equation}
\Delta\tau_{\theta}=\frac{\Delta t}{R_c/c(1-\beta\cos\theta)}
\end{equation}
 Norris et al.(1993, 1996) had pointed the entire
range of all pulse widths is $\sim10 ms-2 s$, that is to say
\begin{equation}
\Delta t\sim10 ms-2 s
\end{equation}
Consequently
\begin{equation}
\frac{\Delta t}{R_c/c}\leq
\Delta\tau_{\theta}\equiv\tau_{\theta,max}-\tau_{\theta,min}
\leq\frac{\Delta t}{R_c/c(1-\beta)}
\end{equation}
where the relation $0\leq\cos\theta\leq1$ has been applied.
Combining (B1), (B5) and (B6), the range of $\Delta\tau_{\theta}$
is limited by
\begin{equation}
10^{-9}\leq\Delta\tau_{\theta}\leq\frac{2}{1000(1-\beta)}
\end{equation}
For one typical value of $\Gamma$, namely $\Gamma$=100, then
$1-\beta=(\Gamma-\sqrt{\Gamma^{2}-1})/\Gamma\simeq 0.00005$ is
decided, therefore
\begin{equation}
10^{-9}\leq\Delta\tau_{\theta}\leq40
\end{equation}
Under this approximation, $\Delta\tau_{\theta}$ is allowed to take
some representative values such as $\Delta\tau_{\theta}$=0.001,
0.01, 0.1, 1, 10 and so on.


\clearpage

\label{lastpage}

\end{document}